\documentstyle[twoside,fleqn,espcrc2,axodraw]{article}

\newcommand{\be}{\begin{equation}}
\newcommand{\ee}{\end{equation}}
\newcommand{\bea}{\begin{eqnarray}}
\newcommand{\eea}{\end{eqnarray}}

\newcommand{\nn}{\nonumber}

\newcommand{\cO}{{\cal O}}

\newcommand{\NPB}[3]{Nucl.\ Phys.\ {\bf B{#1}} (19{#2}) {#3}}
\newcommand{\PRD}[3]{Phys.\ Rev.\ {\bf D{#1}} (19{#2}) {#3}}
\newcommand{\PLB}[3]{Phys.\ Lett.\ {\bf B{#1}} (19{#2}) {#3}}
\newcommand{\PRL}[3]{Phys.\ Rev.\ Lett.\ {\bf {#1}} (19{#2}) {#3}}

\newcommand{\ks}{{\mbox k \!\!\! /}}

\newcommand{\as}{\alpha_{s}}

\newcommand{\AmS}{{\protect\the\textfont2
  A\kern-.1667em\lower.5ex\hbox{M}\kern-.125emS}}

\title{On the pinch technique beyond one loop}

\author{N.J.\ Watson
\address{
Division de Physique Th\'eorique,
IPN,
F-91406 Orsay Cedex, France.}
}
\begin{document}

\begin{abstract}
A brief account is given of the problems involved in
extended the pinch technique beyond the one-loop level,
and of investigations into one possible approach 
to solving them.
\end{abstract}

\maketitle

\section{INTRODUCTION}

The pinch technique (PT) \cite{PT1,PT2} is a well-defined algorithm for the
{\em rearrangement}\, of contributions to conventional one-loop
$n$-point functions in gauge theories to construct one-loop
``effective'' $n$-point functions with 
improved theoretical properties. 
Most notably, the PT one-loop $n$-point functions
are entirely gauge-independent, and satisfy simple QED-like Ward identities.
This rearrangement 
of one-loop perturbation theory 
is based on the
systematic use of the tree level Ward identities 
to cancel among Feynman integrands all factors of longitudinal
four-momentum associated with gauge fields propagating in loops.
As a result of these improved properties, 
the PT has been advocated \cite{PT3,PT4} as the 
appropriate theoretical framework for a wide range of applications 
in which one is forced to go beyond the strictly order-by-order 
computation of $S$-matrix elements, or to consider amplitudes for 
explicitly off-shell processes.

A fundamental criticism of the PT, however, is that
it has yet to be shown how 
the PT algorithm may be consistently extended beyond
the one-loop level to construct 1PI
multi-loop $n$-point functions with the same desirable properties
as at one loop.
This extension requires the solution of two
problems:
 (i) How to reorganize consistently 
contributions to multi-loop integrands
so as to isolate explicitly the 
PT $n$-point
functions as internal loop corrections? 
 (ii) 
How to deal consistently with the factors of longitudinal
internal gauge field four-momentum which themselves originate from
such internal loop corrections?

Here we briefly describe investigations into
one possible approach to solving these problems.

\section{ON THE PT AT TWO LOOPS}

We consider the construction of the PT two-loop
i.e.\ $\cO(\as^{2})$ gluon self-energy in massless QCD,
starting from the four-fermion process
$q\bar{q} \rightarrow q\bar{q}$.
In the class of linear covariant gauges,
the required Feynman rules 
are (cf.\ Fig.~1)
\bea
{\rm Fig.~1a\!}: \!\!\!\!\!\!\!
& &
\frac{1}{Z_{2}}\frac{i\delta_{ij}}{\ks + i\epsilon} 
\label{fig1a} \\
{\rm Fig.~1b}\!: \!\!\!\!\!\!\!
& &
Z_{1}ig\gamma_{\mu}T_{ji}^{m} 
\label{fig1b} \\
{\rm Fig.~1c}\!: \!\!\!\!\!\!\!
& &
\frac{-i\delta^{mn}}{k^{2}\!+\! i\epsilon}\biggl\{ \!
\frac{1}{Z_{3}}\biggl(g_{\mu\nu}\! -\! \frac{k_{\mu}k_{\nu}}{k^{2}}\biggr)
\!+\! \xi\frac{k_{\mu}k_{\nu}}{k^{2}} \!\biggr\}
\label{fig1c} \\
{\rm Fig.~1d}\!: \!\!\!\!\!\!\!
& &
Z_{1,{\rm YM}}gf^{rst}
\Gamma_{\rho\sigma\tau}^{(0)}(k_{1},k_{2},k_{3})
\label{fig1d} \\
{\rm Fig.~1e}\!:\!\!\!\!\!\!\!
& & 
\frac{1}{Z_{2,{\rm FP}}}\frac{i\delta^{rs}}{k^{2} + i\epsilon} 
\label{fig1e} \\
{\rm Fig.~1f}\!:\!\!\!\!\!\!\!
& & 
-Z_{1,{\rm FP}}gf^{rst}k_{2\tau}\,\,.
\label{fig1f}
\eea
The renormalization constants $Z_{i}$ have
perturbative expansions
$Z_{i}
=
1 + \sum_{n=1}^{\infty}
(Z_{i}-1)^{(n)}\alpha_{s}^{n}$.
For order-by-order book keeping, the expressions
(\ref{fig1a})--(\ref{fig1f}) are then split into lowest order
($Z_{i} = 1)$ and associated counterterm contributions.
Here, however, it will be convenient to leave the Feynman
rules as given, with the expansion implicit.

\begin{picture}(200,140)( 0,60)
 

\ArrowLine(15,170)(45,170)
\put( 30,185){\makebox(0,0)[c]{\footnotesize $k$}}
\put(10,170){\makebox(0,0)[c]{\footnotesize $i$}}
\put(50,170){\makebox(0,0)[c]{\footnotesize $j$}}
\put( 30,155){\makebox(0,0)[c]{\small (a)}}

\Gluon(15,120)(35,120){1}{4}
\ArrowLine(35,120)(45,137)
\ArrowLine(45,103)(35,120)
\put(52,145){\makebox(0,0)[c]{\footnotesize $j$}}
\put(52, 95){\makebox(0,0)[c]{\footnotesize $i$}}
\put( 0,120){\makebox(0,0)[c]{\footnotesize $\mu,m$}}
\put( 30, 90){\makebox(0,0)[c]{\small (b)}}

\Gluon(85,170)(115,170){1}{6}
\put(95,177){\vector(1,0){10}}
\put(100,185){\makebox(0,0)[c]{\footnotesize $k$}}
\put(73,170){\makebox(0,0)[c]{\footnotesize $\mu,m$}}
\put(127,170){\makebox(0,0)[c]{\footnotesize $\nu,n$}}
\put(100,155){\makebox(0,0)[c]{\small (c)}}

\Gluon(85,120)(105,120){1}{4}
\Gluon(105,120)(115,137){1}{4}
\Gluon(115,103)(105,120){1}{4}
\put(90,127){\vector(1,0){10}}
\put(109,105){\vector(-1,2){5}}
\put(116,132){\vector(-1,-2){5}}
\put(70,120){\makebox(0,0)[c]{\footnotesize $k_{3},\tau,t$}}
\put(120,145){\makebox(0,0)[c]{\footnotesize $k_{2},\sigma,s$}}
\put(120, 95){\makebox(0,0)[c]{\footnotesize $k_{1},\rho,r$}}
\put(100, 90){\makebox(0,0)[c]{\small (d)}}

\DashArrowLine(155,170)(185,170){2}
\put(170,185){\makebox(0,0)[c]{\footnotesize $k$}}
\put(150,170){\makebox(0,0)[c]{\footnotesize $r$}}
\put(190,170){\makebox(0,0)[c]{\footnotesize $s$}}
\put(170,155){\makebox(0,0)[c]{\small (e)}}

\Gluon(155,120)(175,120){1}{4}
\put(160,127){\vector(1,0){10}}
\DashArrowLine(175,120)(185,137){2}
\DashArrowLine(185,103)(175,120){2}
\put(140,120){\makebox(0,0)[c]{\footnotesize $k_{3},\tau,t$}}
\put(195,145){\makebox(0,0)[c]{\footnotesize $k_{2},s$}}
\put(195, 95){\makebox(0,0)[c]{\footnotesize $k_{1},r$}}
\put(170, 90){\makebox(0,0)[c]{\small (f)}}

\put(-10,70){\makebox(0,0)[l]{\small
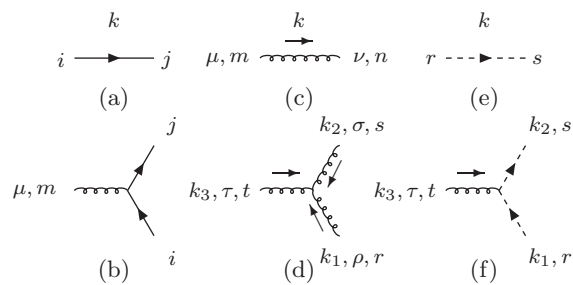
Fig.~1. The diagrams for the Feynman rules
(\ref{fig1a})--(\ref{fig1f}).}}

\end{picture}

\begin{picture}(200,130)(0,80)
 

\Line(  0,170)(  0,200)
\Gluon(  0,185)(15,185){1}{3}
\ArrowArc(25,185)(10,-90,270)
\Gluon( 35,185)(50,185){1}{3}
\Line( 50,170)( 50,200)
\put( 25,160){\makebox(0,0)[c]{\small (a)}}

\Line( 75,170)( 75,200)
\Gluon( 75,185)(90,185){1}{3}
\GlueArc(100,185)(10,  0,180){1}{7}
\GlueArc(100,185)(10,180,360){1}{7}
\Gluon(110,185)(125,185){1}{3}
\Line(125,170)(125,200)
\put(100,160){\makebox(0,0)[c]{\small (b)}}

\Line(150,170)(150,200)
\Gluon(150,185)(165,185){1}{3}
\DashArrowArc(175,185)(10,-90,270){2}
\Gluon(185,185)(200,185){1}{3}
\Line(200,170)(200,200)
\put(175,160){\makebox(0,0)[c]{\small (c)}}

\Line(  0,120)(  0,150)
\Gluon( 0,135)( 40,135){1}{8}
\GlueArc(40,135)(10,-90, 90){1}{7}
\Line( 40,120)( 40,150)
\put( 25,115){\makebox(0,0)[c]{\small + reversed}}
\put( 25,105){\makebox(0,0)[c]{\small (d)}}

\Line( 75,120)( 75,150)
\Gluon( 75,135)(100,135){1}{5}
\Gluon(100,135)(125,145){1}{5}
\Gluon(100,135)(125,125){-1}{5}
\Line(125,120)(125,150)
\put(100,115){\makebox(0,0)[c]{\small + reversed}}
\put(100,105){\makebox(0,0)[c]{\small (e)}}

\Line(150,120)(150,150)
\Gluon(150,145)(200,145){1}{10}
\Gluon(150,125)(200,125){1}{10}
\Line(200,120)(200,150)
\put(175,115){\makebox(0,0)[c]{\small + crossed}}
\put(175,105){\makebox(0,0)[c]{\small (f)}}

\put(-10,90){\makebox(0,0)[l]{\small
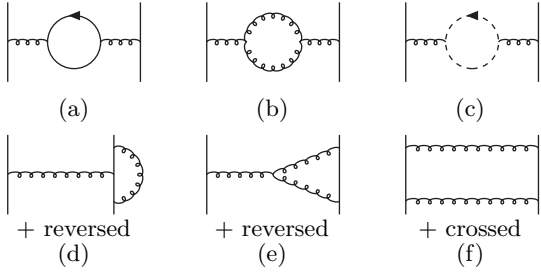
Fig.~2. The one-loop corrections to
$q\bar{q} \rightarrow q\bar{q}$ in QCD.}}

\end{picture}

In the PT at one loop,
the renormalization constants satisfy
the QED-like relation
\be
\frac{Z_{1}}{Z_{2}}
=
\frac{Z_{1,{\rm YM}}}{Z_{3}} 
= 
\frac{Z_{1,{\rm FP}}}{Z_{2,{\rm FP}}}
=
1
\label{ZsST}
\ee
to $\cO(\as)$, and are gauge-independent.

We consider first the $\cO(\as^{2})$ corrections to
$q\bar{q} \rightarrow q\bar{q}$ consisting of
one-loop diagrams with one-loop counterterm insertions
(cf.\ Fig.~2).
Given that each loop contributes a correction of
$\cO(\alpha_{s})$, the $Z_{i}$ must be expanded to
$\cO(\alpha_{s})$.

For the diagram 2a, using (\ref{ZsST}), the Feynman rules for the
quarks are
\bea
{\rm Fig.~1a}\!: \!\!\!\!\!\!\!
& &
\frac{1}{Z_{1}}\frac{i\delta_{ij}}{\ks + i\epsilon} 
\label{fig1aPT} \\
{\rm Fig.~1b}\!: \!\!\!\!\!\!\!
& &
Z_{1}ig\gamma_{\mu}T_{ji}^{m} \,\,.
\label{fig1bPT}
\eea
The above rules hold for $Z_{1}$ to $\cO(\alpha_{s})$.
It follows that, to this order, the renormalization constants 
for the propagators and vertices of the quark loop in
Fig.~2a cancel (alternatively, the four $\cO(\alpha_{s})$
counterterm insertions in the quark loop sum to zero).
This is identical to the case of the two-loop vacuum polarization
in QED.

For the diagrams 2b--2f,
in order to implement the PT we first write
\bea
(\ref{fig1c})
\!\!&=&\!\!
\frac{1}{Z_{3}}\frac{-i\delta^{mn}}{k^{2} + i\epsilon}\biggl(
g_{\mu\nu} - (1-Z_{3}\xi)\frac{k_{\mu}k_{\nu}}{k^{2}}\biggr) \,\,,
\label{fig1c2} \\
(\ref{fig1d})
\!\!&=&\!\!
Z_{1,{\rm YM}}gf^{rst}\Bigl(
\Gamma_{\tau\rho\sigma}^{(0)F}(k_{3};k_{1},k_{2}) \nn \\
& &
+
\Gamma_{\tau\rho\sigma}^{(0)P}(k_{3};k_{1},k_{2}) \Bigr)\,\,,
\label{fig1d2}
\eea
where (\ref{fig1d2}) is the now-familiar decomposition \cite{PT1,PT2}
\bea
\Gamma_{\tau\rho\sigma}^{(0)F}(k_{3};k_{1},k_{2})
&=&
(k_{1}-k_{2})_{\mu}g_{\rho\sigma} \nn \\
& &
-2k_{3\rho}g_{\sigma\mu} +2k_{3\sigma}g_{\rho\mu}\,\,,
\label{Gamma0F} \\
\Gamma_{\tau\rho\sigma}^{(0)P}(k_{3};k_{1},k_{2})
&=&
-k_{1\rho}g_{\sigma\mu} + k_{2\sigma}g_{\rho\mu}\,\,,
\label{Gamma0P}
\eea
for $A_{\tau}^{t}(k_{3})$ the external gluon and
$A_{\rho}^{r}(k_{1})$,
$A_{\sigma}^{s}(k_{2})$ the internal gluons.
The implementation of the PT is then precisely as usual
at one loop, except for
(i) $\xi \rightarrow Z_{3}\xi$ and
(ii) overall factors of $Z_{i}$ for each diagram,
which from (\ref{ZsST}) are the same to $\cO(\alpha_{s})$.
But the PT one-loop $n$-point functions are
individually $\xi$-independent, so (i) is irrelevant.
Thus, implementating the PT as usual, and using (\ref{ZsST}),
the resulting PT rearrangement of the
$\cO(\alpha_{s}^{2})$ contributions to
$q\bar{q} \rightarrow q\bar{q}$ from the diagrams 2b--2f
corresponds to
the Feynman rules
\bea
{\rm Fig.~1c}\!: \!\!\!\!\!\!\!
& &
\frac{1}{Z_{1,{\rm YM}}}
\frac{-i\delta^{mn}}{k^{2} + i\epsilon}g_{\mu\nu} 
\label{fig1cPT} \\
{\rm Fig.~1d}\!: \!\!\!\!\!\!\!
& &
Z_{1,{\rm YM}}gf^{rst}\Gamma_{\tau\rho\sigma}^{(0)F}(k_{3};k_{1},k_{2})
\label{fig1dPT} \\
{\rm Fig.~1e}\!: \!\!\!\!\!\!\!
& & 
\frac{1}{Z_{1,{\rm FP}}}\frac{i\delta^{rs}}{k^{2} + i\epsilon} 
\label{fig1ePT} \\
{\rm Fig.~1f}\!: \!\!\!\!\!\!\!
& & 
-Z_{1,{\rm FP}}gf^{rst}(k_{1} + k_{2})_{\tau}\,\,.
\label{fig1fPT}
\eea
The above rules hold for $Z_{1,{\rm YM}}$,
$Z_{1,{\rm FP}}$ to $\cO(\alpha_{s})$.
Using the above PT Feynman rules for
the self-energy diagrams 2b and 2c,
the renormalization constants $Z_{1,{\rm YM}}$
and $Z_{1,{\rm FP}}$ associated with the gluon and ghost, 
respectively, propagating in the loop cancel
(alternatively, the four $\cO(\alpha_{s})$
counterterm insertions in both the gluon and the ghost loop
sum to zero).
This is exactly analogous to the fermion case above,
and to QED.

In essence, using the gluon Feynman rules 
directly in the form
(\ref{fig1c}), (\ref{fig1d}), 
thence (\ref{fig1c2}), (\ref{fig1d2}), 
we have used not only
the lowest order ($Z_{i} = 1$) components but also the
corresponding $\cO(\as)$ counterterms to trigger the
PT rearrangement. 
This results in the above property
for the spin 0, $\frac{1}{2}$ and 1 contributions to the
$\cO(\as^{2})$ gluon self-energy.

We note that in the background field method (BFM),
the $\cO(\as)$ counterterm insertions in the 
gluon loop in Fig.~2b
do not vanish \cite{BFM} except in the Landau
quantum gauge $\xi_{Q} = 0$. Thus, in the above approach,
the correspondence between the PT
$n$-point functions and those of 
the BFM at $\xi_{Q} = 1$
\cite{BFMPT} does not persist beyond one loop.

We now turn to the $\cO(\as^{2})$ corrections to
$q\bar{q} \rightarrow q\bar{q}$ consisting of two-loop diagrams.
In general, in order
to avoid problems with renormalizability,
we are now obliged to make decompositions 
analogous to (\ref{fig1c2}), (\ref{fig1d2})
for the one-loop corrections to 
the gauge field propagator and 
triple gauge ver-

\begin{picture}(200,60)(-75,100)
 

\Line(  0,120)(  0,150)
\Gluon( 0,142)( 18,142){1}{3}
\ArrowArc(25,142)( 7,-90,270)
\Gluon(32,142)( 50,142){1}{ 3}
\Gluon( 0,128)( 50,128){-1}{10}
\Line( 50,120)( 50,150)
\put( 25,110){\makebox(0,0)[c]{\small (a)}}

\Line( 75,120)( 75,150)
\Gluon( 75,135)( 90,135){1}{3}
\Line( 90,135)(107,145)
\Line( 90,135)(107,125)
\ArrowLine(107,145)(107,125)
\Gluon(107,145)(125,145){1}{4}
\Gluon(107,125)(125,125){-1}{4}
\Line(125,120)(125,150)
\put(100,110){\makebox(0,0)[c]{\small (b)}}

\put( -85,150){\makebox(0,0)[l]{\small
Fig.~3. Some two- }}

\put( -85,140){\makebox(0,0)[l]{\small
loop corrections}}

\put( -85,130){\makebox(0,0)[l]{\small
to $q\bar{q} \rightarrow q\bar{q}$ }}

\put( -85,120){\makebox(0,0)[l]{\small
involving one }}

\put( -85,110){\makebox(0,0)[l]{\small
fermion loop.}}

\end{picture}

\noindent
tex.
This requires the solution of the first problem in Sec.~1. 
For the restricted case of 
the subset of two-loop diagrams for
$q\bar{q} \rightarrow q\bar{q}$ involving always one fermion
loop, this is straightforward, since
the one-loop corrections to the gluon propagator and vertex
are then due only to the quark loops, and so are trivially
isolated.
Furthermore,
the tree level triple gauge vertices which occur
then always have one external leg and two internal legs,
and so can be dealt with as in the PT at one loop.

We thus have to isolate the
factors of longitudinal gluon four-momentum
due to the invariant tensor structure of these
internal fermion loops.

For the diagrams involving internal
quark loop two-point corrections 
$\Pi_{\mu\nu}^{(1,f)}$, e.g. Fig.~3a, we use 
\be
i\Pi_{\mu\nu}^{(1,f)}(k)
=
i(k^{2}g_{\mu\nu} - k_{\mu}k_{\nu})\Pi^{(1,f)}(k^{2})\,.
\label{Pidecomp}
\ee

For the diagrams involving internal quark loop
three-point corrections
$\Gamma_{\rho\sigma\tau}^{(1,f)}(k_{1},k_{2},k_{3})$,
e.g.\ Fig.~3b, 
we use the general decomposition 
\cite{ball}
\bea
\lefteqn{\Gamma_{\rho\sigma\tau}(k_{1},k_{2},k_{3})
\,=\,
 A(k_{1}^{2},k_{2}^{2};k_{3}^{2})g_{\rho\sigma}(k_{1}-k_{2})_{\tau}
 } \nn \\
& &
-B(k_{1}^{2},k_{2}^{2};k_{3}^{2})g_{\rho\sigma}k_{3\tau} \nn \\
& &
-C(k_{1}^{2},k_{2}^{2};k_{3}^{2})
[(k_{1}k_{2})g_{\rho\sigma} - k_{1\sigma}k_{2\rho}]
(k_{1}-k_{2})_{\tau} \nn \\
& & 
+ {\textstyle \frac{1}{3}}S(k_{1}^{2},k_{2}^{2},k_{3}^{2})
(k_{1\tau}k_{2\rho}k_{3\sigma} - k_{1\sigma}k_{2\tau}k_{3\rho}) \nn \\
& &
+F(k_{1}^{2},k_{2}^{2};k_{3}^{2})
[(k_{1}k_{2})g_{\rho\sigma} - k_{1\sigma}k_{2\rho}] \nn \\
& &\times
[k_{1\tau}(k_{2}k_{3}) - k_{2\tau}(k_{1}k_{3})] \nn \\
& &
+H(k_{1}^{2},k_{2}^{2},k_{3}^{2})
\{-g_{\rho\sigma}[k_{1\tau}(k_{2}k_{3}) - k_{2\tau}(k_{1}k_{3})]
\nn \\
& &
+ {\textstyle \frac{1}{3}}
(k_{1\tau}k_{2\rho}k_{3\sigma} - k_{1\sigma}k_{2\tau}k_{3\rho}) \}
\!+\! {\rm cyc.\,prms.}
\label{tgvball}
\eea
Writing $k_{2\rho} = -(k_{1}+k_{3})_{\rho}$ and
$k_{1\sigma} = -(k_{2}+k_{3})_{\sigma}$
in (\ref{tgvball}), 
the extension to arbitrary order of
the decomposition (\ref{fig1d2}) is
\bea
\lefteqn{\Gamma_{\tau\rho\sigma}(k_{3},k_{1},k_{2})\,\,= } 
& & \nn\\
& & 
\Gamma_{\tau\rho\sigma}^{F}(k_{3};k_{1},k_{2}) +
\Gamma_{\tau\rho\sigma}^{P}(k_{3};k_{1},k_{2})
\label{GammaFP}
\eea
where
\bea
\lefteqn{ \Gamma_{\tau\rho\sigma}^{P}(k_{3};k_{1},k_{2}) 
\,\,=\,\,
  P_{\tau\sigma}(k_{1},k_{2})k_{1\rho} }
& & 
\nn \\
& &
- P_{\tau\rho}(k_{2},k_{1})k_{2\sigma} 
+ Q_{\tau}(k_{1},k_{2})k_{1\rho}k_{2\sigma}
\label{PPQ}
\eea
with
\bea
\lefteqn{ P_{\tau\sigma}(k_{1},k_{2})
\,=\,
-[A(k_{2}^{2},k_{3}^{2};k_{1}^{2}) 
+ B(k_{2}^{2},k_{3}^{2};k_{1}^{2})]g_{\tau\sigma} }\nn \\
& &
\,+\,C(k_{1}^{2},k_{2}^{2};k_{3}^{2})(k_{1}-k_{2})_{\tau}k_{3\sigma}
+[C(k_{2}^{2},k_{3}^{2};k_{1}^{2}) \nn \\
& &
- (k_{1}k_{3})F(k_{2}^{2},k_{3}^{2};k_{1}^{2})]
[(k_{2}k_{3})g_{\tau\sigma} - k_{2\tau}k_{3\sigma}] \nn \\
& &
- S(k_{1}^{2},k_{2}^{2},k_{3}^{2})k_{1\tau}k_{3\sigma}
\nn \\
& &
- F(k_{1}^{2},k_{2}^{2};k_{3}^{2}) 
[(k_{2}k_{3})k_{1\tau}-(k_{1}k_{3})k_{2\tau}]k_{3\sigma}  \nn \\
& &
+ H(k_{3}^{2},k_{1}^{2},k_{2}^{2})
[(k_{1}k_{3})g_{\tau\sigma} - k_{1\tau}k_{3\sigma}] \,,
\label{P} \\
\lefteqn{ Q_{\tau}(k_{1},k_{2})
\,=\,
C(k_{1}^{2},k_{2}^{2};k_{3}^{2})(k_{1}-k_{2})_{\tau} } \nn \\
& &
\,-\,F(k_{1}^{2},k_{2}^{2};k_{3}^{2})
[(k_{2}k_{3})k_{1\tau}-(k_{1}k_{3})k_{2\tau}]\,\,.
\label{Q}
\eea

For the one-loop fermionic contribution to
$\Gamma_{\rho\sigma\tau}(k_{1},k_{2},k_{3})$,
the Ward identity
\be
k_{3}^{\tau}\Gamma_{\rho\sigma\tau}^{(1,f)}(k_{1},k_{2},k_{3})
=
\Pi_{\rho\sigma}^{(1,f)}(k_{1}^{2})
\,-\, 
\Pi_{\rho\sigma}^{(1,f)}(k_{2}^{2})
\label{Gamma1fwid} 
\ee
uniquely determines the one-loop fermionic
contributions to the functions $A$, $B$, $C$ and $S$:
\bea
A^{(1,f)}(k_{1}^{2},k_{2}^{2};k_{3}^{2})
\!\!\!\!&=&\!\!\!\!
-\frac{1}{2}\Bigl( \Pi^{(1,f)}(k_{1}^{2}) + \Pi^{(1,f)}(k_{2}^{2}) \Bigr) 
\nn \\
B^{(1,f)}(k_{1}^{2},k_{2}^{2};k_{3}^{2})
\!\!\!\!&=&\!\!\!\!
-\frac{1}{2}\Bigl( \Pi^{(1,f)}(k_{1}^{2}) - \Pi^{(1,f)}(k_{2}^{2}) \Bigr)
\nn \\
C^{(1,f)}(k_{1}^{2},k_{2}^{2};k_{3}^{2})
\!\!\!\!&=&\!\!\!\!
\frac{2}{k_{1}^{2} - k_{2}^{2}} 
B^{(1,f)}(k_{1}^{2},k_{2}^{2};k_{3}^{2})
\nn \\
S^{(1,f)}(k_{1}^{2},k_{2}^{2},k_{3}^{2})
\!\!\!\!&=&\!\!\!
0\,\,. \label{ABCS1f}
\eea
The one-loop massless quark contributions to the functions $F$ and $H$
may be found in \cite{osland}.

Substituting the expressions 
(\ref{ABCS1f})
into (\ref{PPQ})--(\ref{Q}), the Ward identity 
(\ref{Gamma1fwid}) decomposes as
\bea
\lefteqn{k_{3}^{\tau}\Gamma_{\tau\rho\sigma}^{(1,f)F}(k_{3};k_{1},k_{2})
\,\,= } \label{Gamma1fFwid} \\
& &
\Bigl(k_{1}^{2}\,\Pi^{(1,f)}(k_{1}^{2})
\,-\,k_{2}^{2}\,\Pi^{(1,f)}(k_{2}^{2})\Bigr)g_{\rho\sigma}\,\,, \\
\lefteqn{k_{3}^{\tau}\Gamma_{\tau\rho\sigma}^{(1,f)P}(k_{3};k_{1},k_{2})
\,\,=} \nn \\
& &
 k_{2\rho}k_{2\sigma}\,\Pi^{(1,f)}(k_{2}^{2})
\,-\,k_{1\rho}k_{1\sigma}\,\Pi^{(1,f)}(k_{1}^{2})\,\,.
\label{Gamma1fPwid} 
\eea

It is important to point out that, obviously,
the {\em integrands}\,
for the scalar functions
$\Pi^{(1,f)}$ in (\ref{Pidecomp}) and
$A^{(1,f)}$--$H^{(1,f)}$ in (\ref{tgvball})
may be projected out from the {\em integrands}\, for
$\Pi_{\mu\nu}^{(1,f)}$ and
$\Gamma^{(1,f)}_{\rho\sigma\tau}$, respectively:
all rearrangements in the PT take place under the
loop momentum integral sign(s).

Using the above decompositions, the PT can be implemented
in a similar way to the one loop case.
One finds \cite{me}
that the resulting mixed bosonic-fermionic
contribution to the 
two-loop gluon self-energy is 
(i) gauge-independent,
(ii) multiplicatively renormalizable by a local
counterterm and
(iii) has ultra-violet diverence
as specified by the corresponding coefficient
$(\frac{20}{3}C_{A} + 4C_{F})T_{F}n_{f}$
of the two-loop QCD $\beta$-function.
These properties are identical to those of 
the two-loop vacuum polarization in QED.
Furthermore, as a result of the Ward identity
(\ref{Gamma1fFwid}), these contributions to the
two-loop fermion self-energy and gluon-fermion
vertex obey the QED-like Ward identity as
in the PT at one loop.

\section{DISCUSSION AND CONCLUSIONS}

In general, the extension of the PT beyond one loop
requires the solution of the two problems in Sec.~1.
For the case of two-loop QCD $n$-point functions 
which involve always
one fermion loop, the first problem is easily dealt with,
so that one can then investigate the second problem.

Here we have described the approach to this second problem
in which {\em all}\, factors of longitudinal
four-momentum associated with gauge fields propagating
in loops are used to trigger the PT rearrangement.
Such factors arise not only from the lowest order
gauge field propagators and triple gauge vertices,
but also from the invariant tensor structure
of internal loop corrections, as well as the
gauge field propagator and triple gauge vertex counterterms.
It was described how, for the two-loop gluon self-energy
constructed in this approach,
the mixed bosonic-fermionic contribution to this function
displays a set of properties precisely analogous to those
of the two-loop vacuum polarization in QED.

For the purely bosonic i.e.\ gluon and ghost
contributions to the two-loop gluon self-energy
in the PT framwork,
the first problem in Sec.~1 has yet to be solved.
However, this problem has been solved for the
case of the PT two-loop quark self-energy \cite{me}.
A key element of this solution is the decomposition
of the tree level triple gauge vertex
\bea
\lefteqn{ \Gamma_{\rho\sigma\tau}^{(0)}(k_{1},k_{2},k_{3})
\equiv } \nn \\
& &
  \Gamma_{\rho\sigma\tau}^{(0)F}(k_{1};k_{2},k_{3})
+ \Gamma_{\sigma\tau\rho}^{(0)F}(k_{2};k_{3},k_{1})  \nn \\
& &
- \Gamma_{\tau\rho\sigma}^{(0)F}(k_{3};k_{1},k_{2})
- 2\Gamma_{\tau\rho\sigma}^{(0)P}(k_{3};k_{1},k_{2})\,.
\label{GammaFFFP}
\eea

In general, it is as yet unclear whether the approach
described here 
is that which is required to yield PT 
multi-loop ``effective'' $n$-point functions with
all of the desirable properties displayed at one loop.
It may well be that the way to resolve this question
will be to extend the analysis of \cite{PT4} to the two-loop level
to relate the PT
two-loop gauge boson self-energy in a massive theory,
e.g.\ the electroweak Standard Model, 
directly to experimental observables via dispersive
techniques.

\end{document}